\documentclass{elsart}

 \usepackage{graphicx}

\usepackage{amssymb}

\begin{document}

\begin{frontmatter}



\title{Phase transition of triangulated spherical surfaces supported by elastic chains with rigid junctions}


\author{T. Endo}, 
\author{M. Egashira},
\author{S. Obata}, 
\author[label1]{H.  Koibuchi}
\ead{koibuchi@mech.ibaraki-ct.ac.jp}

\address[label1]{Department of Mechanical and Systems Engineering, Ibaraki National College of Technology, 
Nakane 866, Hitachinaka,  Ibaraki 312-8508, Japan}

\begin{abstract}
A surface model with skeletons is investigated by using the canonical Monte Carlo simulations. 
The skeleton is composed of linear chains, which are joined to each other at the rigid junctions. A one-dimensional bending energy is defined on the linear chains, and no two-dimensional curvature energy is assumed on the surface. The model undergoes a first-order transition between the smooth phase and the crumpled phase. We conclude that the first-order transition of the surface model with skeletons is independent of whether the junctions are elastic or rigid.
\end{abstract}

\end{frontmatter}


\section{Introduction}\label{intro}
 A well-known surface model is the one of Helfrich, Polyakov and Kleinert (HPK) \cite{HELFRICH-1973,POLYAKOV-NPB1986,KLEINERT-PLB1986}. The phase structure of HPK model is connected to string models and membranes \cite{WHEATER-JP1994,NELSON-SMMS2004,David-TDQGRS-1989,David-SMMS2004,Wiese-PTCP2000,Bowick-PREP2001,Gompper-Schick-PTC-1994} and therefore has long been investigated theoretically \cite{Peliti-Leibler-PRL1985,DavidGuitter-EPL1988,PKN-PRL1988,BKS-PLA2000,BK-PRB2001} and numerically \cite{KANTOR-NELSON-PRA1987,KANTOR-SMMS2004,WHEATER-NPB1996,BCFTA-JP96-NPB9697,KD-PRE2002,KOIB-PRE-2004-1,KOIB-PRE-2005-1,KOIB-NPB-2006,CATTERALL-NPBSUP1991,AMBJORN-NPB1993,ABGFHHM-PLB1993,BCHHM-NPB9393,KOIB-PLA20023,KOIB-PLA-2004,KOIB-EPJB-2005,KOIB-EPJB-2006}.

 Membranes are considered to be two-dimensional surfaces in the conventional HPK model, and hence homogeneity is assumed in the model. However, membranes are not always homogeneous; cell membranes are known to be supported by skeletons. Hop diffusion of membrane proteins or lipids was experimentally observed recently \cite{Kusumi-BioJ-2004}. The free diffusion of molecules is prohibited, and the diffusion is localized in some domains on the cell membranes. The origin of this is considered to be due to the cytoskeletons. Artificial membranes are considered to have skeletons because they are partly polymerized \cite{CNE-PRL-2006}.

The phase structure of skeleton models should therefore be studied as a statistical mechanical problem. 
Skeleton models for the cytoskeleton were already investigated in \cite{BBD-BioPJ-1998}. A hard-wall and hard-core potential was assumed on the polymer chains with junctions, and the responses to some external stress and the compression modulus were obtained \cite{BBD-BioPJ-1998}. Giant fluid vesicles coated with skeletons was experimentally investigated, and the mechanical properties were reported \cite{HHBRMC-PRL-2001}, where the actin filaments introduce an inhomogeneous structure in homogeneous artificial membranes. A compartmentalized surface model was reported to undergo a first-order transition \cite{KOIB-PLA-2006-2}, where the compartment boundary prevents the vertices from the free diffusion. In \cite{KOIB-JSTP-2006-1}, the phase structure of a surface model with skeleton was investigated, and it was reported that the model has a first-order transition between the smooth phase and the crumpled phase. The interaction of the model in \cite{KOIB-JSTP-2006-1} is described by one-dimensional bending energy for linear chains (or bonds) and two-dimensional bending energy for junctions. The two-dimensional Gaussian bond potential is also assumed in the Hamiltonian. Thus, the interaction of the model in \cite{KOIB-JSTP-2006-1} seems a little bit more complicated than that of HPK model. One of the reasons of this is because the two-dimensional elasticity is assumed at the junctions. 

Therefore, it is interesting to see whether the transition of \cite{KOIB-JSTP-2006-1} occurs in more simplified skeleton models. One possible model is obtained from the model of \cite{KOIB-JSTP-2006-1} by replacing the elastic junction with a rigid junction. 

In this Letter, we study the rigid junction model by using the canonical Monte Carlo (MC) simulations and see how the transition of \cite{KOIB-JSTP-2006-1} occurs in such a simplified model. We must note that the rigid junction model is not included in the elastic junction models. In fact, there are two types of elasticity at the elastic junctions; one is the out-of-plane elasticity and the other is the in-plane elasticity. The former elasticity can be rigid in the elastic junction model of \cite{KOIB-JSTP-2006-1} in the limit of infinite bending rigidity $b_J\!\to\! \infty$, however, the latter one can not be controlled in the elastic junction model. Therefore, the rigid junction model in this Letter and the elastic junction model in  \cite{KOIB-JSTP-2006-1} are considered to be two different models.

\section{Triangulated surfaces}\label{surfaces}
\begin{figure}[htb]
\unitlength 0.1in
\begin{picture}( 0,0)(  10,10)
\put(18,8.5){\makebox(0,0){(a) $(2322,600,42,6)$ }}%
\put(39,8.5){\makebox(0,0){(b) $(6522,1482,42,11)$ }}%
\end{picture}%
\vspace{0.5cm}
\centering
\includegraphics[width=10.5cm]{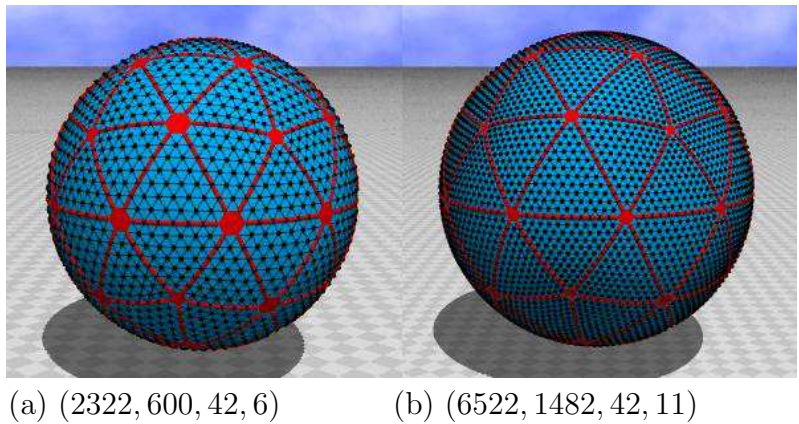}
\caption{Starting configurations of surfaces for MC of size (a) $(N,N_S,N_J,L)\!=\!(2322,600,42,6)$ and  (b) $(N,N_S,N_J,L)\!=\!(6522,1482,42,11)$, where $N$ is the total number of vertices including the total number of junctions, $N_S$ is the total number of vertices on the chains, $N_J$ is the total number of junctions, and $L$ is the length of chains between junctions. Thick lines denote the chains, which terminate at the junctions. } 
\label{fig-1}
\end{figure}

The model is defined on the triangulated surfaces, which are characterized by $N$ the total number of vertices including the junctions, $N_S$ the total number of vertices on the chains, $N_J$ the total number of junctions, and $L$ the length of chains between junctions. The junctions are assumed as rigid plates; twelve of them are pentagon and the others are hexagon. It should be noted again that $N_J$ is included in $N$; a junction is counted as a vertex. Figures \ref{fig-1}(a) and \ref{fig-1}(b) show surfaces of size  $(N,N_S,N_J,L)\!=\!(2322,600,42,6)$ and $(N,N_S,N_J,L)\!=\!(6522,1482,42,11)$, respectively. Thick lines denote the chains terminated at the junctions, and the chains together with junctions form the compartment boundary. All the vertices can fluctuate only locally on the chains as well as inside the compartments, and they are prohibited from the diffusion because of the fixed connectivity nature of the lattice. 

The lattice of $(N,N_S,N_J,L)\!=\!(2322,600,42,6)$ in Fig.\ref{fig-1}(a) corresponds to that of $(N,N_S,N_J,L)\!=\!(2562,600,42,6)$ in Fig.\ref{fig-1}(a) of \cite{KOIB-JSTP-2006-1}. The reason why $N\!=\!2322$ of the surface in Fig.\ref{fig-1}(a) is smaller than $N\!=\!2562$ of the one in \cite{KOIB-JSTP-2006-1} is because the junctions are rigid objects in this Letter while they are elastic ones in \cite{KOIB-JSTP-2006-1}. One hexagonal junction reduces $N$ by $6$, and one pentagonal junction also reduces $N$ by $5$ if they were assumed as rigid objects. Thus, we can check that $2562\!=\! 2322\!+\!6\times 30 \!+\!5\times 12$, where $30$ and $12$ are the total number of pentagonal junction and that of hexagonal junctions, respectively.

The triangulated surfaces such as the ones shown in Figs. \ref{fig-1}(a) and \ref{fig-1}(b)  are constructed as follows: Firstly, we process the icosahedron by dividing every edge into $\ell$-pieces of uniform length, and obtain a triangulated surface of size $N\!=\!10\ell^2\!+\!2$ (= the total number of vertices), where $\ell$ is the number of division of an edge in the icosahedron. The obtained surfaces are thus characterized by $N_5\!=\!12$ and $N_6\!=\!N\!-\!12$, where $N_q$ is the total number of vertices with co-ordination number $q$. Those $N_5(\!=\!12)$-vertices and $N_6(\!=\!N\!-\!12)$-vertices respectively form pentagonal junctions and hexagonal junctions together with their nearest neighbor vertices, in the next stage. Secondly, compartmentalized structures are obtained by dividing $\ell$ further into $m$-pieces ($m\!=\!1,2,\cdots$), and we have uniform chains of length $L\!=\!(\ell /m) \!-\!2$. Those chains terminate at the above-described pentagonal or hexagonal junctions. These junctions include $6$ (or $7$) vertices and $10$ (or $12$) bonds according to whether they are pentagonal or hexagonal, and hence each junction has many degrees of freedom at this stage. Thirdly, the junctions are assumed as rigid objects on the triangulated surfaces: the pentagonal (hexagonal) junction is identified with a regular pentagonal (regular hexagonal) plate. As a consequence, the $6$ (or $7$) vertices together with the  $10$ (or $12$) bonds are identified with a pentagon (or a hexagon).  Thus, we have compartmentalized surfaces with rigid junctions as those shown in Figs. \ref{fig-1}(a) and \ref{fig-1}(b). 

 On the surfaces shown in Figs. \ref{fig-1}(a) and \ref{fig-1}(b), we have $L\!=\!6$, $\ell\!=\!16$ and $L\!=\!11$, $\ell\!=\!26$, respectively. We also note that the compartmentalized structures shown in Figs. \ref{fig-1}(a) and \ref{fig-1}(b) are identical to those of the model in \cite{KOIB-PLA-2006-2} except the fact that the junctions are rigid. The total number $N_C$ of the compartments depends on the surface size $N$. Consequently, $N_C$ is increased with increasing $N$ just as in the surfaces in \cite{KOIB-PLA-2006-2}. We choose the chain length $L$ between the junctions as constant so that it is independent of $N$ in the model. We fix the chain length $L$ such that
\begin{equation}
\label{number-inside}
L=6,\quad L=11,
\end{equation}
 which respectively correspond to the values $n\!=\!21$,  $n\!=\!66$, the total number of vertices inside a compartment \cite{KOIB-PLA-2006-2}. The reason why we fix $n$ is because the size of compartment is considered to be finite in the cell membranes, and also because it is expected that total number of lipids in the compartment remains finite in the cell membranes.
 
\section{Model}\label{model}
The Hamiltonian of the model is given by a linear combination of the two-dimensional Gaussian bond potential $S_1$, the one-dimensional bending energy $S_2$, which are defined by
\begin{equation}
\label{Disc-Eneg} 
S_1=\sum_{(ij)} \left(X_i-X_j\right)^2,\quad S_2=\sum_{(ij)} (1-\cos \theta_{(ij)}). 
\end{equation} 
In these expressions, $\sum_{(ij)}$ in $S_1$ denotes the sum over all the bonds $(ij)$ connecting the vertices $i$ and $j$, and $\sum_{(ij)}$ in $S_2$ denotes the sum over bonds $i$ and $j$, which contain not only bonds in the chains but also {\it virtual bonds} that connect the center of a rigid junction and the neighboring vertices on the chains. The symbol $\theta_{(ij)}$ in $S_2$ is the angle between the bonds $i$ and $j$, which include the virtual bonds described just above. The Gaussian potential $S_1$ is defined not only on the chains but also on all other bonds. As a consequence, the model is considered to be a surface model, although the mechanical strength is maintained by one-dimensional elastic skeletons joined to each other at the rigid junctions. 

\begin{figure}[htb]
\centering
\includegraphics[width=10.5cm]{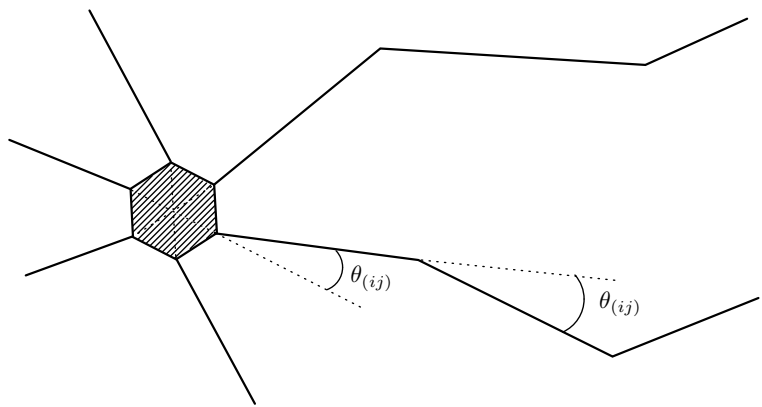}
\caption{A hexagonal junction connected to chains. The angle $\theta_{(ij)}$ in $S_2$ is defined not only at the vertices on the chains but also at the corners (=virtual vertices) of the junction. The triangular lattices attached to the chains were eliminated from the schematic drawing for simplicity. } 
\label{fig-2}
\end{figure}

Figure \ref{fig-2} shows a hexagonal rigid junction connected to chains, where the angle $\theta_{(ij)}$ is defined not only at the vertices on the chains but also at the corners (={\it virtual vertices}) of the junction. The triangular lattices attached to the chains were eliminated from the schematic drawing in Fig. \ref{fig-2} for simplicity.

We must comment on the size of the junctions. The junctions are two-dimensional objects and therefore have their own size to be fixed. The size of junction can be specified by the edge length $R$; the perimeter length of the pentagonal (hexagonal) junction is therefore expressed by $5R$ ($6R$). In this Letter, we fix the size of the junctions such that
\begin{equation} 
\label{junctionsize}
R=0.1.
\end{equation}
The value $R\!=\!0.1$ is quite smaller than that of the elastic junctions in \cite{KOIB-JSTP-2006-1}, where the edge length squared is $R^2\simeq 0.5$ because of the relation $S_1/N\!=\!1.5$ satisfied in the model of \cite{KOIB-JSTP-2006-1}. Since the junctions are two-dimensional rigid objects, it is expected that large-sized junctions influence equilibrium properties of the surface. Therefore, in order to reduce such unclear size-effects we choose the size $R$ relatively small such as in Eq.(\ref{junctionsize}).

We must note that the junctions in the snapshots of Figs. \ref{fig-1}(a),(b) were drawn larger than that expected from Eq.(\ref{junctionsize}); the junction size in the snapshots is fixed according to  $S_1/N\!\simeq\!1.5$. In the following section, we discuss how do we fix the size $R$ to the assumed value in Eq.(\ref{junctionsize}) in the MC simulations.

The partition function $Z$ of the model is defined by
\begin{equation} 
\label{Part-Func}
 Z = \int^\prime \prod _{i=1}^{N} d X_i \exp\left[-S(X)\right],\quad  
 S(X)=S_1 + b S_2, 
\end{equation} 
where $b$ is the bending rigidity corresponding to the one-dimensional bending energy, and  $\int^\prime$ denotes that the center of the surface is fixed. The integration $\prod _{i=1}^{N} d X_i$ is a product of the integration over vertices and that of junctions such that 
\begin{equation} 
\label{integration}
\prod _{i=1}^{N} d X_i = \prod _{{\rm vertices}\; i} d X_i  \prod _{{\rm junctions}\; i} d X_i, 
\end{equation} 
where $\prod _{{\rm junctions}\; i} d X_i$ is the integration over the degrees of freedom for three-dimensional translations and rotations. 

The bending rigidity $b$ has unit of $kT$, where $k$ is the Boltzmann constant, and $T$ is the temperature. The surface tension coefficient $a$ of $S_1$ is fixed to $a\!=\!1$; this is always possible because of the scale invariant property of the model. In fact, in the expression $aS_1 \!+\! b S_2 $ we immediately understand that $a\!=\!1$ is possible, because the factor $a$ of $S_1$ can be eliminated due to the scale invariance of the partition function. Since the unit of $a$ is $(1/{\rm length})^2$, the length unit of the model is given by $\sqrt{1/a}$. We use the unit of length provided by $\sqrt{1/a}\!=\!1$ in this Letter, although $a$ is arbitrarily chosen to be fixed.  

It must also be noted on the relation of the size $R$ in Eq.(\ref{junctionsize}) to the thermodynamic limit $N\!\to\! \infty$. As discussed in the last part of the previous section, the limit $N\!\to\! \infty$ is taken in our model so that the compartment size remains constant as in the model of \cite{KOIB-PLA-2006-2}.  As a consequence, $R$ is also remained constant in the limit $N\!\to\! \infty$. 

\section{Monte Carlo technique}\label{MC-Techniques}
The integration $\prod _{i=1}^{N} d X_i$ in the partition function is done by the canonical Metropolis technique. The update of $X$ in MC can be divided into two steps, which are corresponding to the integrations $\prod _{{\rm junctions}\; i}  d X_i$ and $\prod _{{\rm vertices}\; i} d X_i$ in Eq.(\ref{integration}). The first is the update of $X$ for the vertices including those on the chains: $X$ are shifted so that $X^\prime \!=\! X\!+\!\delta X$, where $\delta X$ is randomly chosen in a small sphere. The new position $X^\prime$ is accepted with the probability ${\rm Min}[1,\exp(-\Delta S)]$, where $\Delta S\!=\! S({\rm new})\!-\!S({\rm old})$. The second is the update of the position of the junctions as three-dimensional rigid objects. This can be further divided into two processes: the first is a random three-dimensional translation, and the second is a random three-dimensional rotation. All of these MC processes are independently performed under about $50\%$ acceptance rate, which is controlled by small numbers fixed at the beginning of the simulations.  We introduce the lower bound $1\times 10^{-8}$ for the area of triangles. No lower bound is imposed on the bond length. 

The junction size $R$ is fixed to $R\!=\!0.1$ in Eq.(\ref{junctionsize}) during the thermalization MCS. The initial value of $R$ is given by $R\!\simeq\! 0.7$ on the surfaces such as those shown in Figs. \ref{fig-1}(a) and \ref{fig-1}(b). Thus, we reduce $R$ from $R\!=\! 0.7$ to $R\!=\!0.1$ by $6\!\times\! 10^{-5}$ at every $25$ MCS in the first $2.5\times 10^6$ MCS. Because of this forced reduction of the junction size, the equilibrium statistical mechanical condition seems to be violated in the first $2.5\times 10^6$ MCS. Therefore, relatively many thermalization ($1.75\times 10^7$ or more) MCS is performed after the reduction. Further thermalization MCS should be performed, if necessary.

We use surfaces of size $(N,N_S,N_J,L)\!=\!(5222,1350,92,6)$, $(9282,2400,162,6)$, $(14502,3750,252,6)$, and $(20882,5400,362,6)$ for the length $L\!=\!6$,  and $(N,N_S,N_J,L)\!=\!(6522,1200,42,11)$, $(14672,2700,92,11)$, and $(26082,4800,162,11)$ for the length $L\!=\!11$. A random number sequence called Mersenne Twister \cite{Matsumoto-Nishimura-1998} is used in the simulations.

\section{Results}\label{Results}
\begin{figure}[htb]
\centering
\vspace{0.7cm}
\unitlength 0.1in
\begin{picture}( 0,0)(  10,10)
\put(19.5,47.5){\makebox(0,0){(a) Collapsed surface at $b\!=\!12.3$ }}%
\put(43,47.5){\makebox(0,0){(b) Smooth surface at $b\!=\!12.4$ }}%
\put(17.2,8.5){\makebox(0,0){(c) The surface section }}%
\put(40,8.5){\makebox(0,0){(d) The surface section}}%
\end{picture}%
\vspace{0.5cm}
\includegraphics[width=10.5cm]{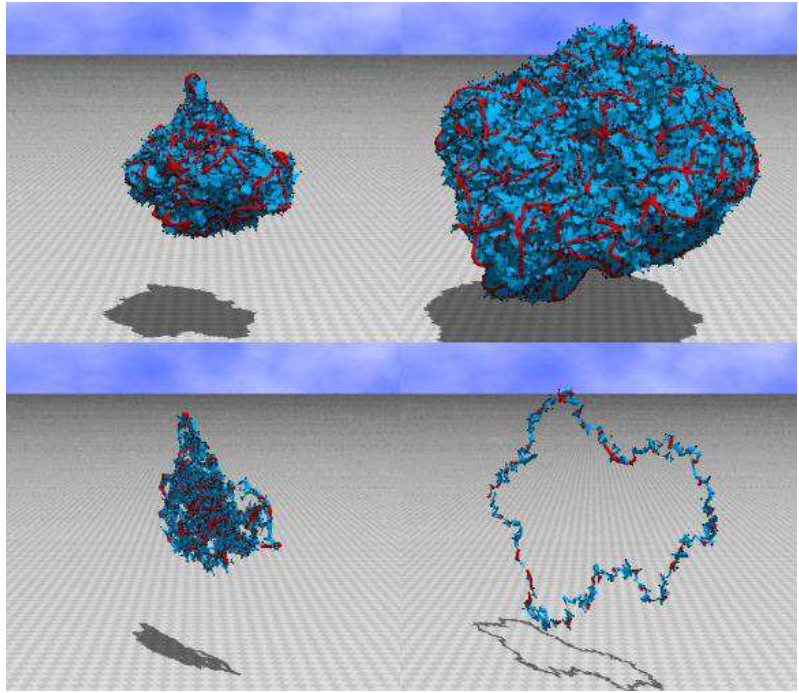}
\caption{Snapshot of the surface of size $(N,N_S,N_J,L)\!=\!(26082,4800,162,11)$ obtained in the crumpled phase at (a) $b\!=\!12.3$ and in the smooth phase at (b) $b\!=\!12.4$, both of which are close to the transition point. The figures are drawn in the same scale.} 
\label{fig-3}
\end{figure}

Snapshots of surfaces are shown in Figs.\ref{fig-3}(a)--\ref{fig-3}(d). Figure \ref{fig-3}(a) is a surface of size $(N,N_S,N_J,L)\!=\!(26082,4800,162,11)$ obtained in the crumpled phase at $b\!=\!12.3$, and Fig.\ref{fig-3}(b) is the one obtained in the smooth phase at  $b\!=\!12.4$. The surface sections of Figs. \ref{fig-3}(a) and \ref{fig-3}(b) are shown in Figs. \ref{fig-3}(c) and \ref{fig-3}(d), respectively. These figures are drawn in the same scale. We immediately see the surface in the smooth phase is actually swollen, while the surface is collapsed in the crumpled phase. 

\begin{figure}[htb]
\centering
\includegraphics[width=10.5cm]{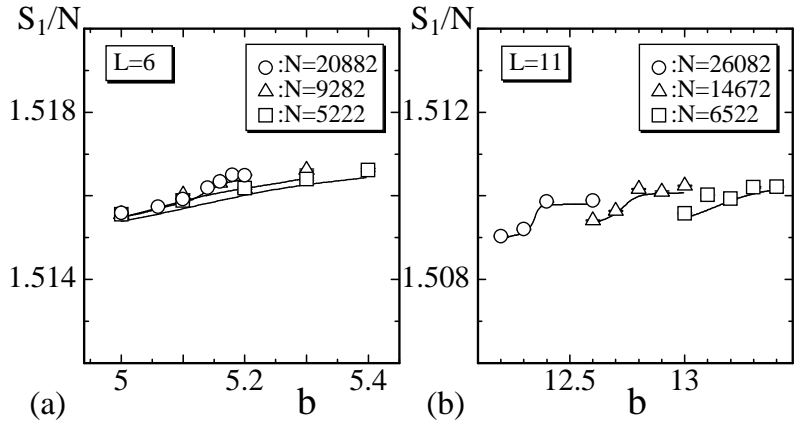}
\caption{The Gaussian bond potential $S_1/N$ vs. $b$ obtained on the surfaces of (a) $L\!=\!6$ and (b) $L\!=\!11$.  $S_1/N$ slightly deviates from $S_1/N\!\simeq\!1.5$. The curves are drawn by the multihistogram reweighting technique.} 
\label{fig-4}
\end{figure}
The Gaussian bond potential $S_1/N$ is shown against $b$ in Figs. \ref{fig-4}(a) and \ref{fig-4}(b), which correspond to the conditions $L\!=\!6$ and  $L\!=\!11$, respectively. The solid curves are drawn by the multihistogram reweighting technique \cite{Janke-histogram-2002}.
 The values of $S_1/N$ in the figures slightly deviate from $S_1/N\!=\!1.5$, which is satisfied on the surface without the rigid junctions or with rigid junctions of negligible size. The reason of this discrepancy is because the surface includes the rigid junctions of finite size. A vertex is the zero-dimensional point, while the rigid junction is the two-dimensional plate and hence shares some area on the surface. We find a gap or a jump in $S_1/N$ of the $(N,N_S,N_J,L)\!=\!(26082,4800,162,11)$ surface in Fig.\ref{fig-4}(b), which can be viewed as a sign of a discontinuous transition.

\begin{figure}[htb]
\centering
\includegraphics[width=10.5cm]{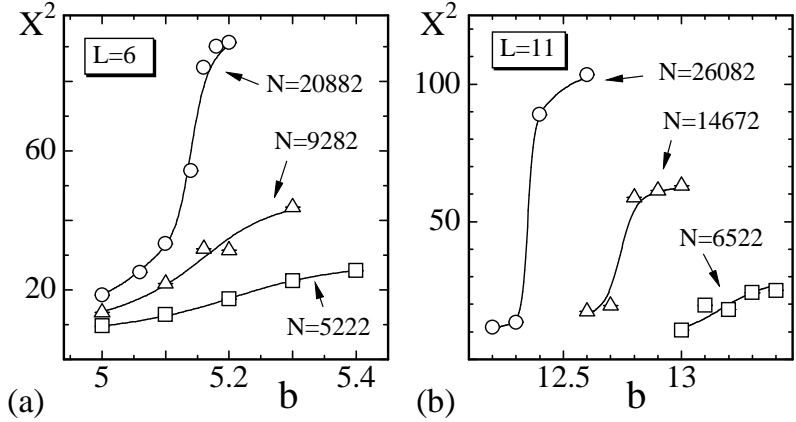}
\caption{The mean square size $X^2$ against $b$ obtained on the surfaces of (a) $L\!=\!6$ and (b) $L\!=\!11$. The curves are drawn by the multihistogram reweighting technique.} 
\label{fig-5}
\end{figure}
The mean square size is defined by
\begin{equation}
\label{X2}
X^2={1\over N} \sum_i \left(X_i-\bar X\right)^2, \quad \bar X={1\over N} \sum_i X_i,
\end{equation}
where $\bar X$ is the center of the surface. The crumpling transition is conventionally understood as the one of surface fluctuations accompanied by surface collapsing phenomena, which can be seen in our surface model; we have seen a collapsed surface and a swollen surface in Figs.\ref{fig-3}(a)--\ref{fig-3}(d). Therefore, we expect that $X^2$ reflects the crumpling transition on spherical surfaces. Figures \ref{fig-5}(a) and \ref{fig-5}(b) are plots of $X^2$ against $b$ obtained under $L\!=\!6$ and  $L\!=\!11$. We find that the variation of $X^2$ becomes sharp against $b$ as $N$ increases. Thus, it is expected that the variation of $X^2$ has a jump at intermediate value of $b$ in either case of $L$. 

The crumpling transition is originally understood as the one for surface fluctuation phenomena. Therefore, the bending energy $S_2/N_S^\prime$ is expected to reflect the transition, where $N_S^\prime$ is given by
\begin{equation}
\label{bendingenegryvertices}
N_S^\prime = N_S + 6N_J-12.
\end{equation}
$S_2/N_S^\prime$ is the bending energy per vertex, because $N_S^\prime$ is the total number of vertices where $S_2$ is defined. $N_S^\prime$ includes virtual vertices which are the corners of the junctions (see also Fig.\ref{fig-2}). Those virtual vertices are not counted as the vertices and hence are not included in $N_S$. Total number of virtual vertices are $6N_J-12$, because the hexagonal junction includes 6-virtual vertices, and the total number of pentagonal junction is $12$. Thus, we have Eq.(\ref{bendingenegryvertices}) for $N_S^\prime$, and therefore we have $N_S^\prime\!=\!1890$, $N_S^\prime\!=\!3360$, $N_S^\prime\!=\!5250$, and $N_S^\prime\!=\!7560$ for the surfaces of length $L\!=\!6$ and size $(N,N_S,N_J,L)\!=\!(5222,1350,92,6)$, $(9282,2400,162,6)$, $(14502,3750,252,6)$, and $(20882,5400,362,6)$;   and $N_S^\prime\!=\!1440$, $N_S^\prime\!=\!3240$, and $N_S^\prime\!=\!5760$ for the surfaces of length $L\!=\!11$ and size $(N,N_S,N_J,L)\!=\!(6522,1200,42,11)$, $(14672,2700,92,11)$, and $(26082,4800,162,11)$.

\begin{figure}[htb]
\centering
\includegraphics[width=10.5cm]{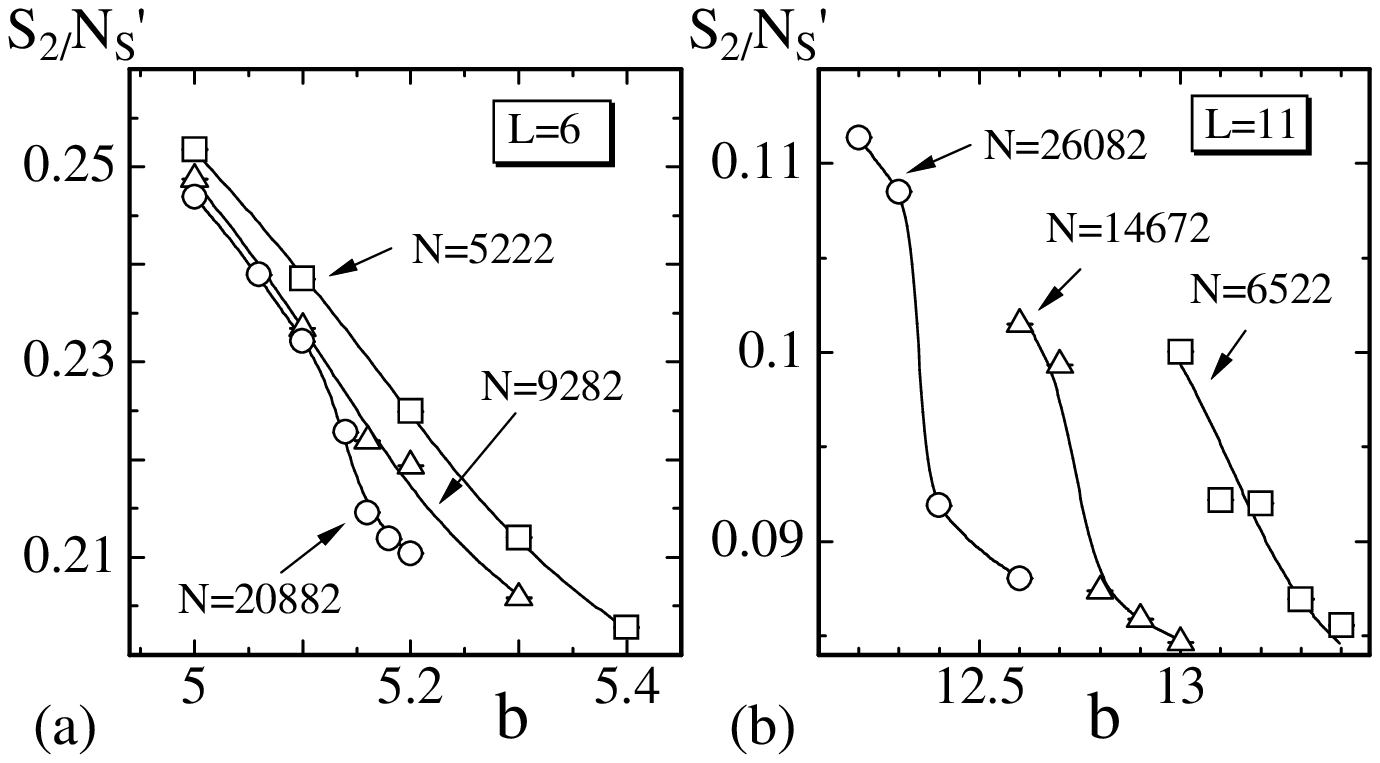}
\caption{The one-dimensional bending energy $S_2/N_S^\prime$ vs. $b$ obtained on the surfaces of (a) $L\!=\!6$ and (b) $L\!=\!11$. } 
\label{fig-6}
\end{figure}

Figures \ref{fig-6}(a) and \ref{fig-6}(b) are plots of $S_2/N_S^\prime$ against $b$ obtained under $L\!=\!6$ and  $L\!=\!11$. We find the expected behavior in $S_2/N_S^\prime$ under both conditions  $L\!=\!6$ and  $L\!=\!11$; $S_2/N_S^\prime$ has a gap (or a jump) at intermediate $b$. This clearly shows that the crumpling transition is of first order.

\begin{figure}[htb]
\centering
\includegraphics[width=10.5cm]{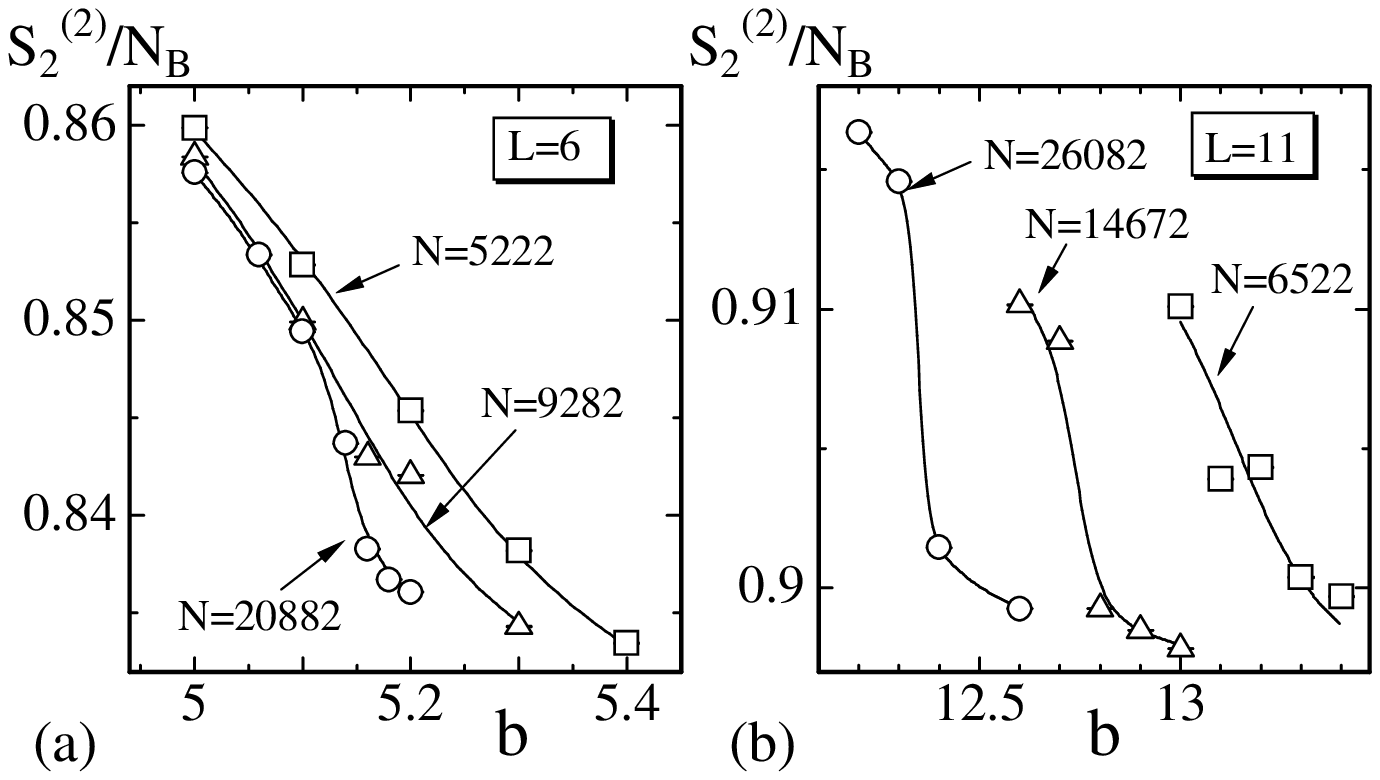}
\caption{The two-dimensional bending energy $S_2^{(2)}/N_B$ against $b$ obtained on the surfaces of (a) $L\!=\!6$ and (b) $L\!=\!11$. $S_2^{(2)}$ is defined by Eq.(\ref{two-dim-bending-energy}) and is not included in the Hamiltonian. $N_B$ is the total number of bonds where $S_2^{(2)}$ is defined.} 
\label{fig-7}
\end{figure}
The transition can also be reflected in the two-dimensional extrinsic curvature, which is defined by
\begin{equation}
\label{two-dim-bending-energy}
 S_2^{(2)}\!=\!\sum_{\langle ij \rangle}(1-{\bf n}_i \cdot {\bf n}_j),
\end{equation}
where ${\bf n}_i$ is the unit normal vector of the triangle $i$. In $S_2^{(2)}$, $\sum_{\langle ij \rangle}$ denotes the summation over all nearest neighbor triangles $i$ and $j$ that have the common bond $\langle ij \rangle$, which includes bonds belonging to the skeleton chains. We denote the total number of bonds, where $S_2^{(2)}$ is defined, by $N_B$, which is given by $N_B\!=\! \sum_{\langle ij \rangle} 1$. Note that $N_B$ includes {\it virtual edges}, which are the one-dimensional edges of the rigid junctions. In fact, we define $S_2^{(2)}$ on the virtual edges, because it is reasonable to define extrinsic curvature on those edges. It is also noted that $S_2^{(2)}$ is not included in the Hamiltonian, and therefore $S_2^{(2)}$ gives no mechanical strength to the surface.

The two-dimensional extrinsic curvature $S_2^{(2)}/N_B$ is plotted in Figs. \ref{fig-7}(a) and \ref{fig-7}(b) against $b$, which are corresponding to the conditions $L\!=\!6$ and  $L\!=\!11$. $N_B$ is the total number of bonds described above. We find that the dependence of $S_2^{(2)}/N_B$ on $b$ shown in Fig.\ref{fig-7} is almost identical to that of $S_2/N_S^\prime$ in Fig.\ref{fig-6}. The gap (or jump) seen in  $S_2^{(2)}/N_B$ also supports that the transition is of first order.

\begin{figure}[htb]
\centering
\includegraphics[width=10.5cm]{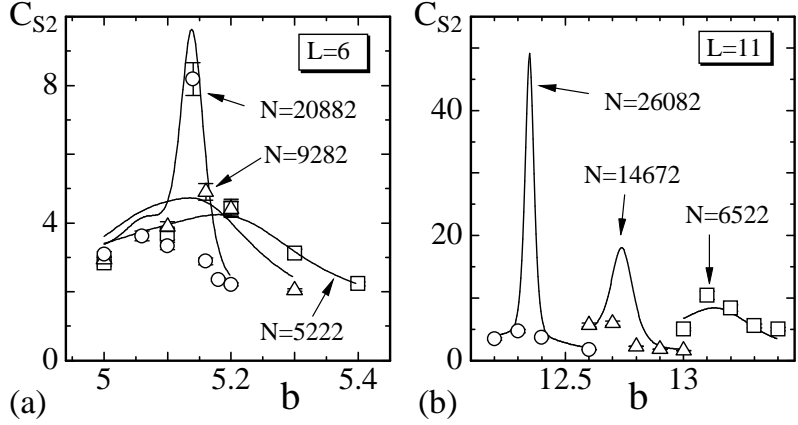}
\caption{The specific heat $C_{S_2}$ for $S_2$ against $b$ obtained on the surfaces of (a) $L\!=\!6$ and (b) $L\!=\!11$. $C_{S_2}$ is defined by Eq.(\ref{specific-heat-1}). The error bars on the symbols are the statistical error, which is obtained by the binning analysis.} 
\label{fig-8}
\end{figure}

The specific heat corresponding to the one-dimensional bending energy $S_2$ is defined by 
\begin{equation}
\label{specific-heat-1}
C_{S_2} \!=\! {b^2\over N_S^\prime} \langle \; \left( S_2 \!-\! \langle S_2 \rangle\right)^2\rangle,
\end{equation}
which can also reflects phase transitions if it has an anomalous behavior. Figures \ref{fig-8}(a) and \ref{fig-8}(b) show $C_{S_2}$ against $b$ obtained under $L\!=\!6$ and  $L\!=\!11$. Solid curves drawn in the figures were obtained by the multihistogram reweighting technique, and those curves clearly show an expected anomalous behavior indicating that $C_{S_2}$ is divergent when $N_S^\prime\to\infty$ (or equivalently $N\to\infty$).   

\begin{figure}[htb]
\centering
\includegraphics[width=10.5cm]{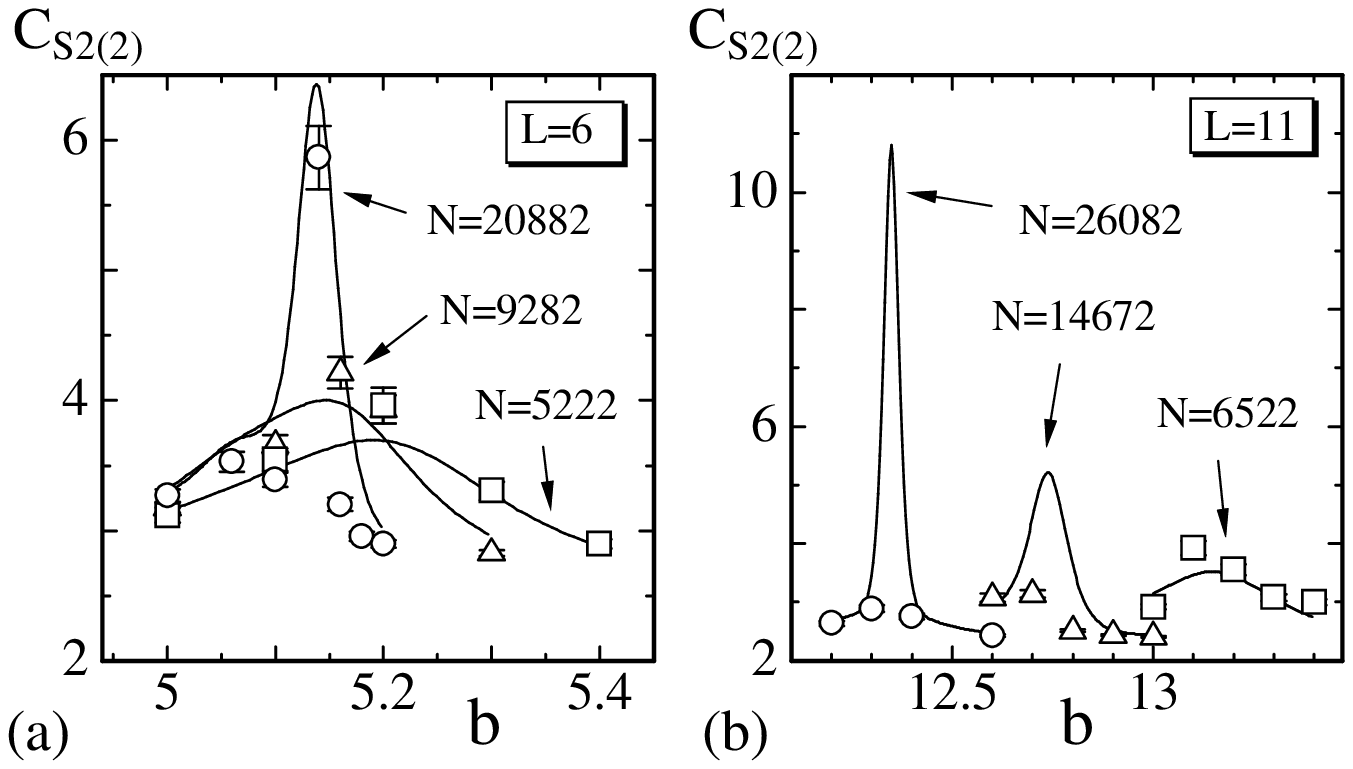}
\caption{The specific heat $C_{S_2^{(2)}}$ for $S_2^{(2)}$ against $b$ obtained on the surfaces of (a) $L\!=\!6$ and (b) $L\!=\!11$. $C_{S_2^{(2)}}$ is defined by Eq.(\ref{specific-heat-2}). The error bars on the symbols are the statistical error, which is obtained also by the binning analysis.} 
\label{fig-9}
\end{figure}
The specific heat corresponding to the extrinsic curvature $S_2^{(2)}$ in Eq.(\ref{two-dim-bending-energy}) can also be defined by 
\begin{equation}
\label{specific-heat-2}
 C_{S_2^{(2)}} \!=\! {1 \over N} \langle \; ( S_2^{(2)} \!-\! \langle S_2^{(2)} \rangle)^2 \rangle, 
\end{equation}
which reflects the transition as $C_{S_2}$ does. Curvature coefficient for $C_{S_2^{(2)}}$ was assumed to be $1$, because $S_2^{(2)}$ is not included in the Hamiltonian and therefore the curvature coefficient is not.  Figures \ref{fig-9}(a) and \ref{fig-9}(b) show $C_{S_2^{(2)}}$ against $b$ obtained under $L\!=\!6$ and  $L\!=\!11$. We can see in $C_{S_2^{(2)}}$ the same anomalous behavior as in $C_{S_2}$. 

\begin{figure}[htb]
\centering
\includegraphics[width=10.5cm]{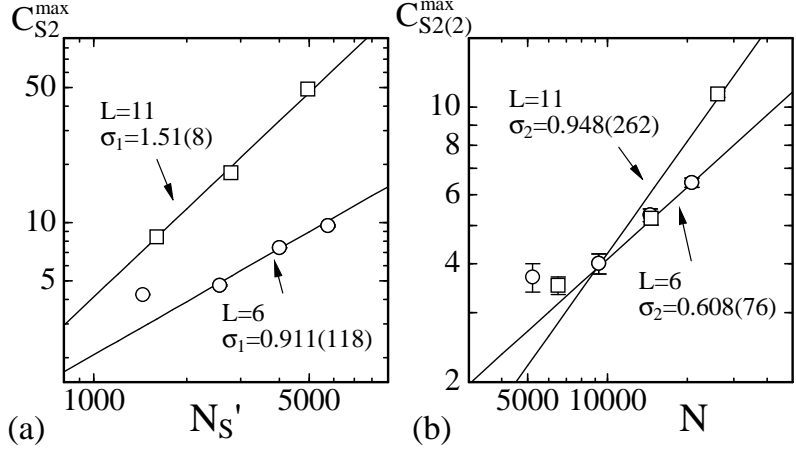}
\caption{ Log-log plots of (a) $C_{S_2}^{\rm max}$ against $N_S^\prime$ and (b) $C_{S_2^{(2)}}^{\rm max}$ against $N^\prime$ obtained on the surfaces of $L\!=\!6$ and $L\!=\!8$. The straight lines are drawn by fitting the largest three data of $C_{S_2}^{\rm max}$ and $C_{S_2^{(2)}}^{\rm max}$ to Eq.(\ref{scaling-exponents}). The peak values and the statistical errors for the fittings were obtained by multihistogram reweighting.} 
\label{fig-10}
\end{figure}
In order to see the anomalous behaviors in $C_{S_2}$ and $C_{S_2^{(2)}}$ in more detail, we plot the peak values of them in Figs. \ref{fig-10}(a) and \ref{fig-10}(b) in log-log scales against $N_S^\prime$ and $N$, respectively. The straight lines were drawn by fitting the data to 
\begin{equation}
\label{scaling-exponents}
C_{S_2}^{\rm max} \propto \left( N_S^{\prime}\right)^{\sigma_1}, \quad C_{S_2^{(2)}}^{\rm max} \propto \left( N \right)^{\sigma_2}, \quad 
\end{equation}
where $\sigma_1$, $\sigma_2$ are critical exponents. Largest three data were used in the fitting in the case $L\!=\!6$. Thus, we have
\begin{eqnarray}
\label{exponents-values}
\sigma_1=0.911\pm 0.118, \quad \sigma_2=0.608\pm 0.076, \quad (L=6), \nonumber \\
\sigma_1=1.51\pm 0.08, \quad \sigma_2=0.948\pm 0.262, \quad (L=11). 
\end{eqnarray}
$\sigma_2\!=\!0.608(76)$ for $L\!=\!6$ is inconsistent with the fact that the transition is of first-order, however, $\sigma_1\!=\!0.911(118)$ is consistent with that. $\sigma_1\!=\!1.51(8)$ and 
 $\sigma_2\!=\!0.948(262)$ under $L\!=\!11$ support the discontinuous transition.

\section{Summary and Conclusion}\label{Conclusion}
A surface model with skeletons has been investigated by the canonical Monte Carlo simulations. The skeletons are composed of one-dimensional linear chains and rigid junctions, whose size is chosen sufficiently small compared to the mean chain-length. The surface is a triangulated sphere and divided into a lot of compartmentalized domains, whose boundary corresponds to the skeletons. The mechanical strength of the surface is given by the skeletons. There is no two-dimensional curvature energy in the Hamiltonian, while one-dimensional bending energy is defined on the chains connected to each other at the junctions. The two-dimensional Gaussian bond potential is included in the Hamiltonian just as in the standard surface model of Helfrich, Polyakov and Kleinert.  The surface model in this Letter is considered to be different from the one with elastic junctions, because the rigid junctions cannot be identified with the elastic junctions due to the property on the in-plane elasticity. 

The surface is characterized by $(N,N_S,N_J,L)$, which are respectively the total number of vertices including the junctions, the total number of vertices on the chains, the total number of junctions, and the length of chains between junctions. The length of chains was fixed to $L\!=\!6$ and $L\!=\!11$, which correspond to $n\!=\!21$ and  $n\!=\!66$ the total number of vertices in a compartment.   

We found that the surface undergoes a first-order crumpling transition between the smooth phase and the crumpled phase. The one-dimensional bending energy $S_2$ has a gap (or a jump) at intermediate bending rigidity $b$, and the two-dimensional extrinsic curvature $S_2^{(2)}$also has a gap at that point. These imply that the surface fluctuations are considered to be a first-order transition. Moreover, it is found that the mean square size $X^2$ also has a gap at the transition point. This implies that the surface-collapsing phenomenon can be viewed as a first-order transition. 

The results in this Letter together with those in \cite{KOIB-JSTP-2006-1} show that the first-order crumpling transition can be seen in the spherical surface model even when the mechanical strength is maintained only by skeletons, which are composed of linear chains joined to each other at the junctions. Moreover, the order of transition is independent of whether the junction is elastic or rigid. 

It has been reported in \cite{KOIB-PLA-2006-2} that the transition can be seen in a compartmentalized surface model, whose mechanical strength is maintained by the two-dimensional curvature of the surface in contrast to the model in this Letter and that in \cite{KOIB-JSTP-2006-1}. Therefore, we can also conclude that the first-order transition occurs in the compartmentalized surface model independently whether the surface is mechanically supported by the skeleton (= the compartment boundary) or by the surface.

Fluidity of lateral diffusion of vertices can be considered by using dynamically triangulated MC technique as in \cite{KOIB-PLA-2006-2}. Vertices freely diffuse inside each compartment on the fluid surfaces supported by the skeletons. It is interesting to see how fluidity influences the transition of the skeleton-supported model. Many interesting problems remain to be studied on the surface model with skeletons.

\section*{Acknowledgment}
This work is supported in part by Grant-in-Aid of Scientific Research, No. 15560160 and No. 18560185. 



\end{document}